\documentclass{llncs}
\usepackage[hyphens]{url}
\usepackage{todonotes}
\usepackage{array}
\usepackage{graphicx}
\usepackage{amssymb}
\usepackage{amsmath}
\usepackage{enumerate}
\usepackage{booktabs}
\usepackage[stable]{footmisc}
\usepackage{bm}

\newcommand{\ZZ}{\mathbb{Z}}
\newcommand{\F}[1]{\mathbb{F}_{#1}}

\newcommand{\fname}{\mathsf}
\newcommand{\aname}[1]{{\fname{#1}}}

\newcommand{\vname}{\mathsf}

\newcommand{\ind}{\approx}
\newcommand{\floor}[1]{\lfloor #1 \rfloor}
\newcommand{\cind}{\underset{C}{\ind}}
\newcommand{\psamplefrom}{\xleftarrow{\vname{PRNG}(H, s)}}
\newcommand{\ceil}[1]{\lceil #1 \rceil}
\newcommand{\samplefrom}{\xleftarrow{\$}}

\newenvironment{keywords}{
       \list{}{\advance\topsep by0.35cm\relax\small
       \leftmargin=1cm
       \labelwidth=0.35cm
       \listparindent=0.35cm
       \itemindent\listparindent
       \rightmargin\leftmargin}\item[\hskip\labelsep
                                     \bfseries Keywords:]}
     {\endlist}

\begin{document}

\title{Lightweight Self-Bootstrapping Multiparty Computations of Time-Series Data with Custom Collusion Tolerance}
\author{Michael Clear\footnotemark[1], Constantinos Patsakis, and Paul Laird}
\institute{School of Computer Science and Statistics,\\Trinity College
Dublin}
\footnotetext[1]{The author's work is funded by the Irish Research Council EMBARK Initiative.}

\maketitle
\begin{abstract}
In this work we compare two recent multiparty computation (MPC) protocols for private summation in terms of performance. Both protocols allow multiple rounds of aggregation from the same set of public keys generated by parties in an initial stage. We instantiate the protocols with a fast elliptic curve and provide an experimental comparison of their performance for different phases of the protocol. Furthermore, we introduce a technique that allows the computational load of both protocols to be reduced at the expense of protection against collusion tolerance. We prove that both protocols remain secure with this technique, and evaluate its impact on collusion tolerance and the number of rounds supported.

\end{abstract}

\section{Introduction}
Modern computing has reached a point that allows users from every part of the globe to exchange information seamlessly. Nevertheless, the environment cannot be considered friendly. Cyber attacks each year are reaching a new peak, while recently disclosed events clearly indicate that privacy measures are not properly deployed or applied. The hostile environment to which modern users are exposed to has triggered the generation of many security and privacy protocols. In many scenarios, users have to co-operate to perform several tasks which result in some of their data being partially or fully exposed. Since the latter might not be an acceptable option, given the sensitivity of the submitted information, there has been much research effort in the area of secure multi-party computation (MPC). 

While there has been much progress made in recent years towards making MPC practical, the efficiency of many protocols is still not acceptable for many real-world applications. To cover this gap, apart from general MPC protocols (which allow any function to be computed), many MPC protocols for specialized computations have been introduced. While they might support only a small set of functions, these protocols can be obtain much improved performance, even in devices with limited computing resources. Thus, apart from the interest in general MPC protocols, many application needs are pushing towards the development of more targeted MPC protocols. A typical example of this trend are protocols which allow privacy-aware summation of values. These protocols are widely used for load monitoring of smart meters, privacy-aware participatory sensing, and generally for submitting time series data with privacy.

Recently, a new protocol was proposed by Patsakis, Clear and Laird \cite{PCL:2014} (referred to here as PCL) that allows privacy-aware multiparty aggregation of values. A notable feature is that it supports multiple rounds of aggregation from the same set of public keys published by the users, and it relies only on the standard Decisional Diffie-Hellman (DDH) assumption.  PCL is based on a protocol due to Kursawe, Danezis and Kohlweiss \cite{kursawe2011privacy} (referred to here as KDK) which also has a variant that supports multiple rounds of aggregation by exploiting bilinear pairings. Both these papers do not consider the performance of their protocols in practice. In fact, \cite{PCL:2014} does not address its performance relative to multi-round KDK. In this paper, we extensively compare the performance of both protocols, and show that in practice PCL outperforms multi-round KDK by a significant margin, largely due to the cost of pairings, but also due to the larger finite field in which recovery of the sum takes place. Following on from concrete performance results, we propose a new extension that is applicable to both protocols, which leads to considerable performance improvements. In PCL, protection against collusion tolerance is traded off against the number of rounds supported. Our extension trades collusion tolerance further (in both protocols) to achieve a performance gain. However, since a tolerance of $1/3 - 1$ of the parties is satisfactory for many real-world applications (note that this corresponds to the Byzantine optimum), we argue that reducing tolerance to (say) $1/3$ is justified given the significant gains in performance.
\subsection{Main contributions}
The main technical contribution of this work are as follows. Firstly, a thorough performance analysis is given of the multi-round protocols PCL and KDK. To the best of our knowledge, this is the first assessment of the practicality of multi-round KDK. Secondly, we introduce a new technique that is applicable to both protocols that allows all parties to reduce their computational load by adjusting their collusion tolerance. We prove security for the extended versions of both protocols that employ this technique.

\subsection{Organization of this work} 
The rest of this work is organized as follows. In the next section we provide a brief overview of the related work in this area, along with a description of the two protocols, PCL and KDK in their single-round and multi round versions. Afterwards, in section 3, we compare the performance of the two protocols. Sections 4 and five describe methods to reduce the computational load of PCL and KDK protocols respectively. 

\section{Related work}
In many application scenarios users have to co-operate to achieve specific tasks, however, they need to know that the that they will send will not be disclosed. Based on this problem, Yao introduced the concept of secure multiparty computation was
\cite{yao1982protocols}. Despite the huge initial advances \cite{yao1986generate,goldreich1987play,chaum1988multiparty,ben1988completeness}, only recently did real-world implementations become practical \cite{malkhi2004fairplay,ben2008fairplaymp,bogetoft2008multiparty}.

Due to efficiency, secure multiparty computation can be categorized into schemes that allow arbitrary computations to be performed without leaking information, and more specialized protocols which allow the private evaluation of particular functions, such as summation. In the first category we have schemes based on Garbled Circuits (efficient implementations include \cite{lindell2007efficient,lindell2012secure,lindell2013fast}), Nielsen's protocol using Oblivious Transfer \cite{nielsen2012new} and Oblivious RAM \cite{Goldreich:1987o,Gordon:2012}. Other protocols use arithmetic circuits such as those based on the BGW protocol e.g: VIFF \cite{Damgaard:2009} and SPDZ which employs fully homomorphic encryption \cite{cryptoeprint:2011:535}. As already discussed, the second category includes more application-specific protocols. For instance for calculating the private sum of $n$ users we have the scheme of  Clifton {\it et al.} \cite{clifton2002tools}, the two round scheme of  Yang {\it et al.} \cite{Yang:2005} and the single round scheme of Shi {\it et al.} but relies on the previously distribution of shares of shares of zeros from a trusted third party \cite{Shi:2011}. Other protocols are focused on calculating one bit multi-party computations, such as DC-net \cite{Chaum88thedining,golle2004dining} and \cite{hao20092}.

\section{Preliminaries}\label{sec:mpc}
\subsection{Security Definition}
We adopt the standard simulation-based definition of security in the semi-honest model.
We base our definition below on Definition 2.1 in \cite{AsharovL11}. Here we consider only computational security, and relax the more standard definition to deterministic functionalities with a single output.  Note that this definition is general enough to accommodate multi-round aggregation.

Let $\vec{m} \in (\{0, 1\}^\ast)^n$ be a vector of the inputs from each party and let $\pi$ be a protocol.  We define the view of a party $P_i$ in the execution of protocol $\pi$ with input vector $\vec{m}$ as 
\[\aname{VIEW}^\pi_i(\vec{x}) = (m_i, r_i, \mu_i^{(1)}, \hdots, \mu_i^{(\ell)})\] where $m_i$ is party $P_i's$ input, $r_i$ is its random coins and $\mu_i^{(1)}, \hdots, \mu_i^{(\ell)}$ are the $\ell$ protocol messages it received during the protocol execution. Similarly, the combined view of a set of $I \subseteq \{1, \hdots, n\}$ parties is denoted by $\aname{VIEW}^\pi_I(\vec{x})$.
\begin{definition}[$t$-privacy of $n$-party protocols for deterministic aggregation functionalities] Let $f : (\{0, 1\}^\ast)^n \to (\{0, 1\}^\ast)$ be a deterministic $n$-ary functionality and let $\pi$ be a protocol that \emph{correctly} computes $f$. We say that $\pi$ if $t$-private if for every $\vec{m} \in (\{0, 1\}^\ast)^n$ where $|m_1| = \hdots = |m_n|$, there exists a PPT algorithm $\mathcal{S}$ such that for every $I \subset \{1, \hdots, n\}$ with $|I| \leq t$, and every $\vec{m} \in (\{0, 1\})^n$ where $|m_1| = \hdots = |m_n|$, it holds that:
\begin{equation}\label{tpriv_sec}
\{\aname{VIEW}^\pi_I(\vec{m})\} \cind \{\mathcal{S}(I, \vec{m}_I, f_I(\vec{m}))\}.
\end{equation}
where $\cind$ denotes computational indistinguishability.
\end{definition}

\subsection{KDK Single-Round Protocol}
Kursawe, Danezis and Kohlweiss (KDK) \cite{kursawe2011privacy} present a specialized multiparty computation (MPC) protocol for private summation, which is shown is be secure in the semihonest model under the Decisional Diffie-Hellman (DDH) assumption. We refer to this protocol as KDK. In their protocol, $n$ parties $P_1, \hdots, P_n$ can compute a joint sum of their inputs $m_1, \hdots, m_n \in \{0, \hdots, \beta\}$ for some positive integer $\beta$. An overview of their protocol follows.

Let $p$ be a prime. The ``public parameters'' used in the protocol consist of a description of a cyclic group $\mathbb{G}$ of order $p$ together with a generator $g$ of $\mathbb{G}$. It is assumed that DDH is intractable in $\mathbb{G}$. These public parameters $\vname{PP} = (\mathbb{G}, g, p)$ are known to all parties $P_i$. The group operation of $\mathbb{G}$ is written multiplicatively.

\begin{enumerate}
\item
\textbf{Setup:} Party $P_i$ generates a secret key $x_i \in \ZZ_p$ and computes her public key $u_i = g^{x_i} \in \mathbb{G}$. She broadcasts $u_i$.
\item
For every $r \in \{1, \hdots, \ell\}$:

\textbf{Main Round:}
\begin{itemize}
\item
Party $P_i$ chooses her input $m_i \in \{0, \hdots, \beta\}$.
\item
Compute $w \gets \prod_{j \in 1}^{i - 1} u_j^{-1} \cdot \prod_{j \in i + 1}^{n} u_j \in G$.
\item
Compute $v_i \gets w^{x_i} \cdot g^{m_i} \in G$.
\item
Broadcast $v_i$.
\end{itemize}
\item
\textbf{Output:}
The protocol produces an output in $\{0, \hdots, n\beta\}$, namely the sum of the user inputs. To compute the sum $\sigma$:
\begin{itemize}
\item
Compute $z \gets \prod_{j = 1}^n v_j$.
\item
Use Pollard's Lambda algorithm to compute the discrete log $\sigma \in \{0, \hdots, n\beta\}$ of $z$ with respect to $g$ in $G$. The time complexity of Pollard's lambda algorithm is $\sqrt{n\beta}$.
\item
Output $\sigma$.
\end{itemize}
\end{enumerate}
It can be easily observed that
\begin{equation}\label{eq:result_aggr}
\prod_{j = 1}^n v_j = g^{\sum_{j = 1}^n m_j}
\end{equation}

\subsection{KDK Multi-Round Protocol}
If the protocol must be run a number of times, it would be desirable to avoid re-running the ``Setup'' phase above which involves each party generating and broadcasting a new public key; in practice, a verification step for these keys may also be needed. To re-use the published keys $u_i, \hdots, u_n$ for more than a single round of aggregation, Kursawe et al. propose an extension of their protocol that facilitates multiple-rounds. In fact, their multi-round protocol accommodates an unbounded number of rounds. They make use of bilinear pairings to achieve this. More details on bilinear pairings are provided in Section \ref{sec:perf} when we address practical issues. We give a very brief overview here that is sufficient to understand the multi-round protocol. The following is based on the definition from \cite{DuttaBS:2004} (Section 2).
\begin{definition}\label{def:pairing}
Let $\mathbb{G}_1$, $\mathbb{G}_2$ and $\mathbb{G}_T$ be cyclic groups of prime order $q$. We write $\mathbb{G}_1$ and $\mathbb{G}_2$ additively, and $\mathbb{G}_T$ multiplicatively. A bilinear pairing $e : \mathbb{G}_1 \times \mathbb{G}_2 \to \mathbb{G}_T$ is an efficiently computable map satisfying
\begin{itemize}
\item
\emph{Bilinearity:} $e(aP, bQ) = e(P, Q)^{ab}$ for all $P \in \mathbb{G}_1$, $Q \in \mathbb{G}_2$ and $a, b \in \ZZ_q^\ast$.
\item
\emph{Non-degeneracy:} If $P$ is a generator for $\mathbb{G}_1$ and $Q$ is a generator for $\mathbb{G}_2$, then $e(P, Q) \neq 1$.
\end{itemize}
\end{definition}
Examples of bilinear pairings (or their modifications) that are used in cryptography include the Tate Pairing \cite{BarretoKLS02,GalbraithHS02}, Weil Pairing \cite{BF03}, and Ate Pairing \cite{HessSV06}.

Let $\mathbb{G}_1$, $\mathbb{G}_2$ and $\mathbb{G}_T$ be cyclic groups of prime order $p$. Let $e : \mathbb{G}_1 \times \mathbb{G}_2 \to \mathbb{G}_T$ be a cryptographic bilinear pairing meeting the conditions of Definition \ref{def:pairing}. Furthermore, the Bilinear Decisional Diffie Hellman (BDDH) assumption is expected to hold with respect to $\mathbb{G}_1$, $\mathbb{G}_2$, $\mathbb{G}_T$ and $e$. Let $H_2 : \ZZ \to \mathbb{G}_2$ be a hash function. The main changes to KDK to support multiple rounds are as follows (optimizations are discussed later):
\begin{itemize}
\item
The public parameters include generators $P \in \mathbb{G}_1$, $Q \in \mathbb{G}_2$ and $g = e(P, Q) \in \mathbb{G}_T$.
\item
The public keys are generated as $U_i \gets x_i P \in \mathbb{G}_1$ for all $1 \leq i \leq n$.
\item
In round $k$, party $P_i$ computes
\begin{itemize}
\item
$Q_k \gets H_2(k) \in \mathbb{G}_2$ (i.e. for a good choice of $H_2$, we have $Q_k = rQ$ for some uniformly random $r$, which is intractable to find).
\item
$w \gets \prod_{j \in 1}^{i - 1} e(U_j, Q_k)^{-1} \cdot \prod_{j \in i + 1}^{n} e(U_j, Q_k) \in \mathbb{G}_T$.
\end{itemize}
\end{itemize}
The rest of the protocol remains unchanged except that the computations are performed in $\mathbb{G}_T$, and $P_i$ may choose a different input value in every round. Naturally, the output of the protocol is then $(\sigma_1, \hdots, \sigma_\ell) \in \{0, \hdots, n\beta\}^\ell$ if $\ell$ rounds are executed.

\subsection{PCL Multi-Round Protocol}
Patsakis, Clear and Laird \cite{PCL:2014} introduced another multi-round variant of KDK without pairings. Their protocol (referred to here as PCL) allows a bounded number of rounds $\ell$ to be performed from the same public key information. However, $\ell$ depends on the acceptable collusion tolerance $t \leq n$. Both single-round and multi-round KDK are $t$-private for any $t \leq n$. On the other hand, in order for PCL to be $t$-private, at most $\ell = \floor{\frac{n - t}{2}}$ rounds are permitted. Concretely, for $t = n/3$ (Byzantine tolerance) and $n = 100$, we can securely run 33 rounds before re-keying. One of the advantages of PCL is that it only relies on the DDH assumption in some cyclic group $\mathbb{G}$ of prime order $p$, like single-round KDK. However, there are also benefits regarding performance over multi-round KDK as highlighted in Section \ref{sec:perf}.

It is observed in \cite{PCL:2014} that KDK is centered on a fixed matrix $A$ with entries in $\{-1, 0, 1\}$ that determine the exponents used to compute $w$. In other words, party $P_i$ raises $P_j$'s public key $u_j$ to the power $A_{i, j}$ when computing $w$. However, the matrix $A$ used in KDK has the form: $A_{i, i} = 0$ for $1 \leq i \leq 0$; $A_{i, j} = 1$ for $1 \leq i < j \leq n$ ($+1$ in upper triangle) and $A_{i, j} = -1$ for $1 \leq j < i \leq n$ ($-1$ in lower triangle). Therefore, $A$ is skew-symmetric i.e. $-A = A^T$. The main idea in PCL is to generate a new skew-symmetric matrix $A^{(k)}$ in a deterministic manner for each round $k$. Furthermore, the matrix $A^{(k)}$ is chosen to have coefficients in $\ZZ_p$ instead of $\{-1, 0, 1\}$ in order to prove security. We refer the reader to \cite{PCL:2014} for details. Here we assume a function $\chi : \ZZ_p \times \ZZ \to \ZZ_p^{n \times n}$ that takes a random seed and a round number, and outputs a pseudorandom s
 kew-symmetric matrix over $\ZZ_p$. Note that the seed can be pre-determined or derived from the users' public keys. The main differences to single-round KDK are as follows:
\begin{itemize}
\item
Let $s \in \ZZ_p$ be a seed deterministically derived form $u_1, \hdots, u_n$.
\item
In round $k$, party $P_i$ computes
\begin{itemize}
\item
$A^{(k)} \gets \chi(s, k) \in \ZZ_p^{n \times n}$.
\item
$w \gets \prod_{j \in 1}^{n} u_j^{A^{(k)}_{i, j}} \in \ZZ_p$.
\end{itemize}
\end{itemize}
The remaining steps are the same as single-round KDK with the exception that each party may choose a different input value in every round, and the final output is $\ell$ values in $\{0, \hdots, n\beta\}$.

\section{Comparing the performance of multi-round KDK and PCL}\label{sec:perf}
We begin by comparing the \emph{original} multi-round KDK from \cite{kursawe2011privacy} and \emph{original} PCL protocol form \cite{PCL:2014}. We are unaware of any concrete performance results for multi-round KDK, which we believe are important in order to assess its practicality. In this section, the performance of both multi-round protocols is measured and compared. The results motivate our proposed optimizations.

\subsection{Computation of a round}
Firstly, we compare the necessary group operations that a party $P_i$ must perform in a given round.  Multi-round KDK requires $n - 1$ pairings, $n - 1$ multiplications in $\mathbb{G}_T$ and an exponentiation in $\mathbb{G}_T$. Note the omission of the inversions in $\mathbb{G}_T$ for $1 \leq j < i$. The reason for this is that in the \emph{Setup} phase, party $P_i$ can compute $U_j \gets -U_j$ for $1 \leq j < i$ where $U_j \in \mathbb{G}_1$ is $P_j$'s public key. Thus by bilinearity of $e$, no inversions are needed in $\mathbb{G}_T$.

On the other hand, PCL needs $n$ exponentiations and $n$ multiplications in group $\mathbb{G}$, Derivation of the \emph{per-round information} for KDK involves computing $Q_k \gets H_2(k) \in \mathbb{G}_2$ whereas PCL involves computing $A^{(k)} \gets \chi(s, k)$. However, the latter can be optimized since only a single row of the matrix $A^{(k)}$ is needed by party $P_i$. As pointed out in \cite{PCL:2014}, $\chi$ uses a hash function $H : \ZZ_p \times \ZZ \times \ZZ \times \ZZ \to \ZZ_p$ to generate $A^{(k)}_{i, j}$; that is, $A^{(k)}_{i, j} \gets H(s, k, i, j)$. However, derivation of the \emph{per-round information} in both protocols is negligible relative to the cost of the group operations.

At present, all known efficient realizations of bilinear pairings are based on elliptic curves. Therefore, in order to implement multi-round KDK, we had to instantiate $\mathbb{G}_1$ and $\mathbb{G}_2$ with elliptic curve groups. There is far less freedom when choosing an elliptic curve when pairings are involved, since the chosen curve must satisfy additional properties. Notwithstanding, to provide a fair performance comparison between both protocols, the same curve was used for both protocols.

Consider an elliptic curve $E$ over $\F{q}$ for prime $q$ whose order is $\#E(\F{q}) = p$. For PCL, the group $\mathbb{G}$ may be instantiated by the additive group formed by $E(\F{q})$. For multi-round KDK, we have opted to use the Modified Tate Pairing to instantiate $e$ since efficient implementations exist in libraries such as MIRACL.

Now the embedding degree $k$ of $E$ is the smallest positive integer such that $p \mid q^k - 1$. Concretely, the Tate pairing takes two points on $E(\F{q^k})[p]$ and outputs an element of $\F{q^k}^\ast$ (more precisely, an element of a multiplicative subgroup of order $p$ of $\F{q_k}^\ast$), where $E(\F{q^k})[p]$ denotes the set of points on $E(\F{q^k})$ of order $p$ i.e. the set of points $A$ with $pA = \mathcal{O}$, where $\mathcal{O}$ is the additive identity (\emph{point at infinity}). Basically, $\mathbb{G}_1$ and $\mathbb{G}_2$ must be two distinct subgroups of $E(\F{q^k})$ of order $p$. In fact, we can set $\mathbb{G}_1$ to $E(\F{q})$ to make the pairing calculation faster. Furthermore, certain pairing-friendly curves $E$ allow us to choose $\mathbb{G}_2$ such that it is isomorphic to a subgroup of $E^\prime(\F{q^{k/d}})$ where $E^\prime$ is related to $E$ (known as the ``twisted curve''); this means arithmetic operations can be carried out in the smaller field $\F{q^{
 k/d}}$ where $d$ is the ``twist degree''.

The curve chosen for our implementation is a member of the pairing-friendly BN family from \cite{Pereira:2011} with a 254-bit prime $q$, embedding degree $k = 12$, and ``twist degree'' $d = 6$. As a consequence of the latter, arithmetic operations in $\mathbb{G}_2$ can be carried out in $\F{q^2}$ instead of $\F{q^{12}}$. In addition, $\mathbb{G}_T$ is the group generated by $g = e(P, Q) \in \F{q^k}^\ast$, and thus its arithmetic operations are carried out in the ``big'' field $\F{q^k}$. This has implications for message recovery, since Pollard's lambda algorithm is much slower.

We implemented both protocols in C++ using the MIRACL C/C++ library version 5\footnote{On Github, commit \url{https://github.com/CertiVox/MIRACL/commit/6d7bb13285e7962ccfa110b4149fa8a63db2ed52}} using the BN curve as described above. The code was compiled with g++ with the compiler flags ``-m64 -O2'' as recommended in the MIRACL documentation, and was executed on a machine with an Intel Core i5-3340M CPU (2.7 GHz) and 4GB of RAM, and running 64-bit Debian GNU/Linux 3.2.41. For each protocol. we measured the time taken to compute a single round per participant (recall that a round involves preparing the value $v_i$ for some party $P_i$). We ran this 100 times for different values of $n$. Note that on each run a random index $i \in \{1, \hdots, n\}$ was chosen, and the round was executed for $P_i$. Our results are shown in Table \ref{tab:round_times}. As expected, the cost of a round is roughly linear in $n$. Moreover, the difference in times between KDK and PCL is significant;
  on average PCL outperforms KDK by a factor of $\approx 437$ based on the times in Table \ref{tab:round_times}). For even a moderate number of users such as $n = 100$, it is clear that multi-round KDK is not suited to time-sensitive applications. This is more pronounced for resource-constrained devices such as wireless sensors.

\begin{table}\label{tab:round_times}
\begin{center}
\begin{tabular}{lrrr}
%\toprule
 & \textbf{n=10} & \textbf{n=100} & \textbf{n=1000} \tabularnewline
\midrule
\textbf{Multi-round KDK} & 47.78 (0.25) & 480.71 (2.34) & 4795.33 (6.69) \tabularnewline

\textbf{PCL} & 0.94 (0.053) & 1.33 (0.01) & 5.33 (0.07) \tabularnewline
\bottomrule
\end{tabular}
\label{results}
\caption{Mean time in ms (over 100 runs) for a party to compute a round (standard deviation in parenthesis).}
\end{center}
\end{table}

\subsection{Recovery of the sum with Pollard's Lambda Algorithm}
Now we turn our attention to the aggregation phase of the protocol. In any given round, this entails multiplying all elements $v_i$ to calculate $z \gets \prod v_i$, then finding a discrete logarithm with respect to a generator $g$ to recover the sum $\sigma = \sum m_i$ of the parties' inputs in the round. For this purpose, Pollard's Lambda algorithm is employed. In PCL, the $v_i$ belong to $\mathbb{G}$ whereas in multi-round KDK, they belong to $\mathbb{G}_T$. Recall that our implementation with elliptic curves instantiates $\mathbb{G}$ as $E(\F{q})$ whereas $\mathbb{G}_T$ is instantiated as a subgroup of $\F{q^k}^\ast$. As such, this phase is more expensive for multi-round KDK because the field operations take place in a ``bigger'' field. Pollard's Lambda algorithm dominates recovery of the sum. Its time complexity is $O(\sqrt{M})$ where $M$ denotes the size of the message space. In this case, $M = n\beta$ since each party chooses her message in $\{0, \hdots, \beta\}$.

\begin{figure}[tbp]
  \begin{center}
    \includegraphics[scale=0.75]{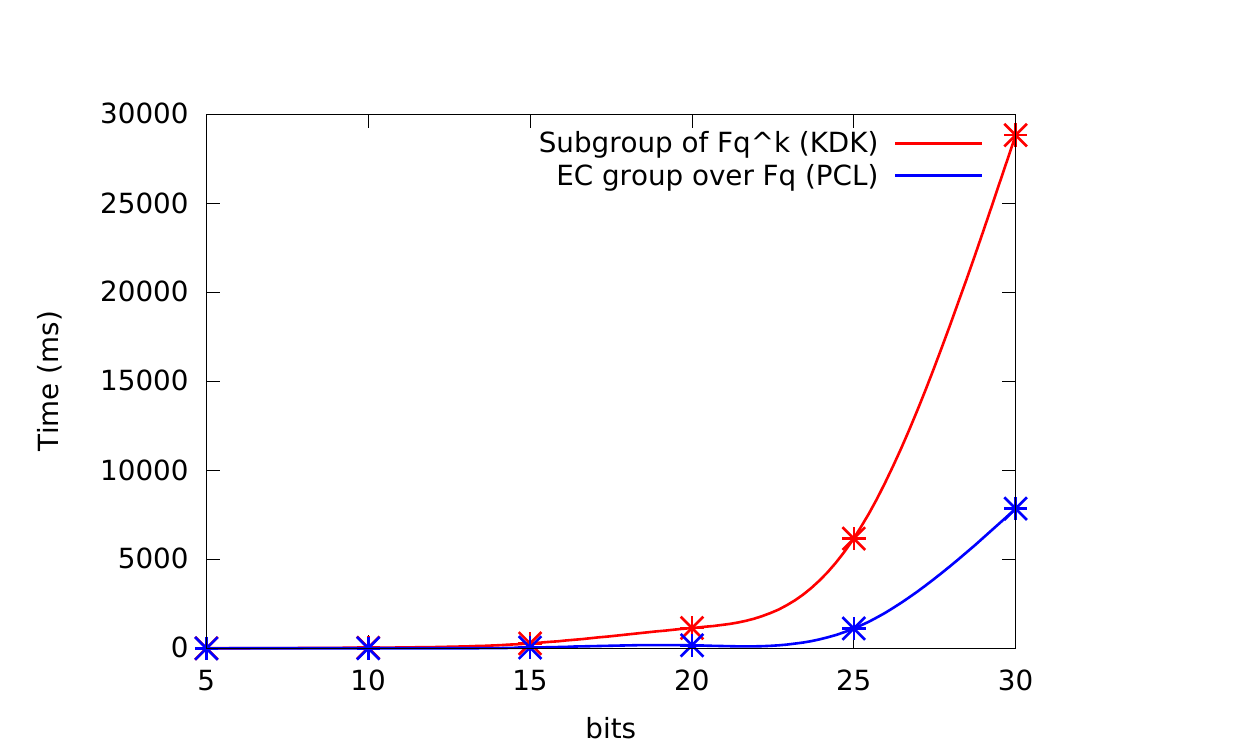}
    \caption{Average time to find discrete logs in $E(\F{q})$ (PCL) and $\F{q^k}$ (multi-round KDK) for different value ranges (upper bound in bits).}
    \label{fig:dl_times}
  \end{center}
\end{figure}

In order to compare multi-round KDK to PCL in this phase, we measured the time taken to compute Pollard's Lambda algorithm for different message space sizes. Moreover, values were randomly generated in the set $m \samplefrom \{2^{b - 2}, \hdots, 2^b\}$ for different values of $b$ and the time taken to recover $m$ given $g^m$ using Pollards Lambda algorithm was measured (the range given to the algorithm was $\{0, \hdots, 2^b\})$; here $g$ denotes the generator of the group in question and multiplicative notation is employed arbitrarily. The measurements were performed in Sage version 5.9 on the same machine and operating system as that used for the previous experiment above. We ran the experiment 10 times each for $b \in \{5, 10, 15, 20, 25, 30\}$ for both $E(\F{q})$ and the group $\langle e(P, Q) \rangle \subset \F{q^k}^\ast$ (recall that $k = 12$ for the curve we used). Our results are shown in Figure \ref{fig:dl_times}. Observe that for $\approx 30$-bit numbers, multi-round
  KDK takes almost half a minute to recover the result. Hence, for large values of $\beta$, it is the recovery phase that acts as the main bottlekneck in multi-round KDK.

\section{Reducing Computational Load in the PCL Protocol}\label{sec:pcl_holes}
Although the performance results from Table \ref{tab:round_times} are favorable for PCL, there is still a motivation for seeking improvements, because many applications involve running the protocol on low-powered devices such as mobile phones. In particular, when a user's phone has low battery, it would be desirable to reduce their computational load. We introduce an optimization that allows parties to adjust their computational workloads at the expense of reducing their protection against collusion tolerance.  Consider a skew-symmetric matrix $A$ generated in a particular round of the protocol. Ordinarily, each row of the matrix has $n - 1$ non-zero entries with overwhelming probability. It is easy to see that the cost of computing $v^{(k)}_i$ for some round $k$ is linear in the number of non-zero entries in the associated matrix $A$. A user $P_i$ can reduce this cost by setting some entries in row $i$ of $A$ to zero. Suppose he sets $\alpha$ entries of the $i$-th row of $A$
  to zero; the zero at $A_{i, i}$ is not included. His computation time 
 is now $\frac{n - 1 - \alpha}{n - 1}$ of the original. For convenience, we refer to these zeros as \emph{holes}. If $P_i$ sets a \emph{hole} at position $j$, then $P_j$ necessarily has a hole at position $i$. So if one party sets $\alpha$ holes, another $\alpha$ parties enjoy a marginally reduced cost. This idea raises some natural questions including
\begin{enumerate}
\item
Since the matrix $A$ is deterministically and non-interactively generated in each round, how are a party's holes set?
\item
What impact does the number of holes $\alpha$ have on the collusion tolerance of the protocol? Does this lead to fewer rounds, and how many?
\item
For some collusion tolerance $t$ and some desired number of rounds $m$, how many users $h$ can simultaneously set $\alpha$ holes while maintaining privacy?
\end{enumerate}
The first question can be addressed by assuming that all parties who seek to place holes declare this intention in the initial stage of the protocol. We define the predicate $\mathfrak{h} : \{1, \hdots, n\} \to \{0, 1\}$ such that for all $1 \leq i \leq n$, we have $\mathfrak{h}(i) = 1$ if and only if $P_i$ sets holes. For all parties $P_i$ with $\mathfrak{h}(i) = 1$, we assume they all set $\alpha$ holes. The following algorithm is used to compute $A$ in a given round. Note that we are using a hash function $H$ with seed $s$ effectively as a PRNG\footnote{A cryptographically secure pseudorandom number generator is not used directly because the proof of security is in the random oracle model, which relies on $H$ being modelled as a random oracle.}. As such, we use the notation $x \psamplefrom X$ to denote the fact that $x$ is pseudorandomly sampled from the set $X$ using a pseduorandom function derived from $H$ and $s$.
\begin{enumerate}
\item
Set $A$ to the $n \times n$ zero matrix.
\item
Create an array $W$ of length $n$; set $W[i] \gets \alpha$ for every $1 \leq i \leq n$.
\item
For all $1 \leq i \leq n$:
\begin{enumerate}
\item
If $\mathfrak{h}(i) = 0$, set $S \gets \{i + 1, \hdots, n\}$.
\item
Else;
\begin{enumerate}
\item
Set $\alpha^\prime \gets W[i]$.
\item
Choose a subset of $\{i + 1, \hdots, n\}$ of cardinality $\alpha^\prime$ (these correspond to the holes). Formally, sample $\hat{S} \psamplefrom \{X \subseteq \{i + 1, \hdots, n\} : |X| = \alpha^\prime\}$.
\item
Set $W[j] \gets W[j] - 1$ for every $j \in \hat{S}$.
\item
Set $S \gets \{i + 1, \hdots, n\} \setminus \hat{S}$.
\end{enumerate}
\item
For each $j \in S$:
\begin{enumerate}
\item
Set $A_{i, j} \psamplefrom \ZZ_p$.
\item
set $A_{j, i} \gets -A^{i, j}$
\end{enumerate}
\end{enumerate}
\end{enumerate}
It turns out the all parties can set $\alpha$ holes and benefit from a performance boost. The following lemma gives a \emph{lower bound} on the number of non-holes a party must set, and therefore the number of exponentiations in $\mathbb{G}$ (viewed multiplicatively) that have to performed per round (per party) to maintain $t$-privacy in the worst-case (this is the number of non-holes plus 1).  Recall that $n$ exponentiations are needed per round in the original PCL protocol.
\begin{lemma}\label{lem:expon}
Let $\ell$ be the number of rounds. Let $t < n$ be the collusion tolerance. 
A lower bound on the number of non-holes a party must set  per round to guarantee $t$-privacy is $2\ell + t$.
\end{lemma}
\begin{proof}
To derive this lower bound, we consider the worst-case scenario. This corresponds to the case where all parties choose the same $\alpha$ positions, which all lie outside the $t$ parties controlled by the adversary.

A skew-symmetric matrix $A \in \ZZ^{n \times n}$ can be viewed as $n$ quadratic equations in $n$ variables $x_1, \hdots, x_n$. Moreover, the $i$-th row of $A$ represents the equation $\sum_{j = 1}^n A_{i, j} x_i x_j$. There are $n(n - 1)/2$ unique monomials $x_i x_j$. There is a corresponding coefficient matrix $A^\prime = \fname{coeff}(A) \in \ZZ^{n \times n(n - 1) / 2}$ for $A$ whose $i$-th row consists of the coefficients for each monomial. By construction, $A^\prime$ is linearly dependant since all $n$ equations must sum to zero to achieve correctness. But, if we remove one row (say the $n$-th) to yield $A^{\prime\prime}$, then $A^{\prime\prime}$ should be linearly independent. This is necessary to ensure security (see the proof of Theorem 1 in \cite{PCL:2014} for more details). If we have $\ell$ rounds, then the resulting coefficient matrix $C \in \ZZ_p^{\ell (n - 1) \times (n - 1) / 2}$  (formed by vertically concatenating the first $n - 1$ rows of $\fname{coeff}A^{(k)}$ for $k \leq \ell$) must be linearly independent to guarantee security. But by dropping the adversary's $t$ parties, along with $\alpha$ parties corresponding to the holes (in the worst-case), we are left with $n - t - \alpha$ parties. A precondition for linear independence is that $\ell \leq \frac{n - t- \alpha}{2}$. Since each party must set $t + ((n - t) - \alpha) + 1$ non-holes, it follows from the inequality that a lower bound is $2\ell + t$.
\qed
\end{proof}
Lemma \ref{lem:expon} only considers a lower bound. This tell us the best we can hope for. We explain in the next section why setting the number of non-holes to merely meet this lower bound is not sufficient for $t$-privacy.

\subsection{Partitioning of the graph}\label{sec:graph}
Let $A$ be a skew-symmetric matrix.  $A$ can be viewed as an undirected graph $G$, where the vertices represent the parties $P_1, \hdots, P_n$, and there is an edge between $P_i$ and $P_j$ if $A_{i, j} \neq 0$. If $G$ can be partitioned into more than one connected component, say components $G^\prime_1$ and $G^\prime_2$, then partial sums can be learned of the parties in both components. It turns out that by just setting non-holes to merely meet the lower bound given by Lemma \ref{lem:expon} is not sufficient to avoid partitions. Since $A$ is deterministically generated, it might be tempting to modify the algorithm to generate $A$ such that this does not occur. However, we don't know ahead of time which $t$ parties are controlled by the adversary. Let $m$ the number of non-holes set by each player. From Lemma \ref{lem:expon}, $m \geq 2\ell + t$. Formally, to ensure $t$-privacy, given a connected graph $G$ of degree $m$, it must hold that if any $\kappa \leq t$ vertices are removed to yield $G^\prime$, then $G^\prime$ remains connected. In fact, such as graph $G$ is referred to as $k$-vertex-connected, where $k$ in this case is $t + 1$. There are techniques to generate $G$ to satisfy this property. One of those techniques involve each party linking to its $m$ ``nearest neighbors'', where the distance between $P_i$ and $P_j$ is $|j - i|$. Therefore, we can change the algorithm that generates the skew-matrix $A$ to follow this technique. This means that we can ensure $t$-privacy and perform only $2\ell + t + 1$ exponentiations per round, as opposed to $n$ in the original protocol.

\section{Reducing Computational Load in the KDK Protocol}\label{sec:kdk_holes}
Due to its comparitively poor performance, as shown in Table \ref{tab:round_times}, there is abundant motivation for reducing the computational cost of a round of multi-round KDK. Our technique from Section \ref{sec:pcl_holes} can also be applied to multi-round KDK, although with even greater scope for improvement. The reason for this is as follows. Recall the coefficient matrix from the proof of Lemma \ref{lem:expon}. In PCL, $n - 1$ rows are added to this matrix in every round. However, in  multi-round KDK, due to the pairing, there is no linear relationship between the set of equations of each round. As such, we only have to consider a single set of $n - 1$ equations in isolation. Since there are fewer equations, it is easier to avoid linear dependence, and more holes can be set as a consequence. Since a skew-symmetric matrix $A$ is fixed for multi-round KDK, we replace this matrix with one generated according to the technique mentioned in Section \ref{sec:graph}. 

Like above, we consider the worst-case scenario and derive a lower bound on $\alpha$.
Let us represent the number of dishonest users as a fraction $\tau \in [0, 1]$ of $n$.
As a necessary condition for linear independence, the following inequality must be satisfied
\begin{equation}\label{eq:ineq_alpha}
((1 - \tau)n - \alpha)((1 - \tau)n - \alpha) \geq 2((1 - \tau)n - 1).
\end{equation}
Since we know that $\alpha < (1 - \tau)n$, we can simplify (\ref{eq:ineq_alpha}) to obtain an upper bound on $\alpha$:
\begin{equation}\label{eq:bound_alpha}
\alpha \leq (\tau - 1)n - 1/2\sqrt{8n - 7} - 1/2.
\end{equation}
Because there will always be $\tau n$ non-holes, the number of additional non-holes that is necessary is $\ceil{1/2\sqrt{8n - 7} + 1/2}$. It follows that the computational load as a fraction of the original load is then lower bounded by
\begin{equation}\label{eq:load_kdk_holes}
\frac{\tau\cdot n + \ceil{1/2\sqrt{8n - 7} + 1/2}}{n} = \tau + \frac{\ceil{1/2\sqrt{8n - 7} + 1/2}}{n}
\end{equation}
which shows that as $n$ grows, the cost of this modified protocol relative to the original converges towards $\tau$.

\section{Conclusions}
While there are several protocols that enable users to privately compute the summary of their values, in many cases, as in the case of KDK, there are several hidden implementation bottlenecks which can significantly delay the calculations of different stages of the protocol. In this work we indicate that for instance the multi-round KDK protocol, due its heavy reliance on pairings is not practical at present for large numbers of users $n > 100$, whereas PCL is highly practical in these cases at the expense of reduced collusion tolerance. We showed that further customization of the privacy level facilitates further extensions to both protocols, and such extensions were shown to be secure. If applied, these extension can boost the efficiency of both these protocols, leading to faster applications with customizable levels of privacy.
\bibliographystyle{plain}
\small{\bibliography{refs}}

\end{document}